\documentclass[twocolumn,amsmath,amssymb,floatfix,pre,floatfix]{revtex4-1}

\usepackage{graphicx,color}
\definecolor{brown}{rgb}{0.63,0.27,0.18}
\definecolor{orange}{rgb}{1.00,0.65,0.00}

\usepackage{afterpage} 

\usepackage{moresize}
\usepackage{dcolumn}
\usepackage{bm,multirow}

\makeatletter
\newcommand*{\balancecolsandclearpage}{%
  \close@column@grid
  \twocolumngrid
}
\makeatother

\newcommand{\be}{\begin{equation}}
\newcommand{\ee}{\end{equation}}
\usepackage{color}

\begin{document}

\newcommand {\rsq}[1]{\left< R^2 (#1)\right.}
\newcommand {\rsqL}{\left< R^2 (L) \right>}
\newcommand {\rsqbp}{\left< R^2 (N_{bp}) \right>}
\newcommand {\Nbp}{N_{bp}}
\newcommand {\etal}{{\em et al.}}
\newcommand{\Ham}{{\cal H}}
\newcommand{\AngeloComment}[1]{\textcolor{red}{(AR) #1}}
\newcommand{\AndreaComment}[1]{\textcolor{blue}{A: #1}}
\newcommand{\JanComment}[1]{\textcolor{green}{(JS) #1}}

\newcommand{\NewText}[1]{\textcolor{orange}{#1}}
\newcommand{\scs}{\ssmall}

\newcommand{\Tau}{\mathrm{T}}



\title{
Topological analysis and the recovery of entanglements in polymer melts
} 

\begin{abstract}
The viscous flow of polymer chains in dense melts is dominated by topological constraints whenever the single chain contour length, $N$, becomes larger than the characteristic scale $N_e$, 
defining comprehensively the macroscopic rheological properties of the highly entangled polymer systems.
Even though the latter are naturally connected to the presence of hard constraints like knots and links within the polymer chains, 
the difficulty of integrating the rigorous language of mathematical topology with the physics of polymer melts has limited somehow a genuine topological approach to the problem of classifying these constraints and to how they are related to the rheological entanglements.
In this work, we tackle this problem by studying the occurrence of knots and links in lattice melts of randomly knotted and randomly concatenated ring polymers of various bending stiffness.
Specifically, by introducing an algorithm which shrinks the chains to their minimal shapes which do not violate topological constraints and by analyzing those in terms of suitable topological invariants, we provide a detailed characterization of the topological properties at the intra-chain level (knots) and of links between pairs and triplets of distinct chains. 
Then, by employing the Z1-algorithm on the minimal conformations in order to extract the entanglement length $N_e$, we show that the ratio $N/N_e$, the number of entanglements per chain, can be remarkably well reconstructed in terms of 2-chain links solely. 
\end{abstract}

\author{Mattia Alberto Ubertini}
\email{mubertin@sissa.it}
\affiliation{Scuola Internazionale Superiore di Studi Avanzati (SISSA), Via Bonomea 265, 34136 Trieste, Italy}

\author{Angelo Rosa}
\email{anrosa@sissa.it}
\affiliation{Scuola Internazionale Superiore di Studi Avanzati (SISSA), Via Bonomea 265, 34136 Trieste, Italy}

\date{\today}




\maketitle

\section{Introduction}\label{sec:Intro}
The viscoelastic behavior of concentrated solutions or melts of {\it linear} polymer chains can be understood assuming~\cite{DeGennesBook,DoiEdwardsBook,RubinsteinColbyBook} slow reptative flow of each chain through the network of topological obstacles (entanglements) formed by the surrounding chains.
According to this picture, entanglements confine each chain within an effective tube-like region of diameter $d_T \approx \langle b \rangle \, n_K \sqrt{N_e / n_K}$
where
$\langle b \rangle$ is the mean bond length,
$n_K$ is the Kuhn length of the polymers (in monomer units~\cite{OnContourLengthUnitsNote}) accounting for the fiber stiffness
while 
the {\it topological} entanglement length $N_e$ is the characteristic, material-dependent~\cite{KavassalisNoolandiPRL,SvaneborgEveraers:2020-1,SvaneborgEveraers:2020-2}, length scale marking the crossover from {\it non-entangled} to {\it entangled} polymer behavior.
Then, the mean size or gyration radius $\langle R_g\rangle$ of polymer chains with contour length $N \gtrsim N_e$ follows the power-law behavior
\begin{equation}\label{eq:LinChainsRg}
\langle R_g \rangle \sim d_T \left( \frac{N}{N_e} \right)^{1/2} \sim \langle b\rangle \, n_K \left( \frac{N}{n_K} \right)^{1/2} \, ,
\end{equation}
and all the essential structural and dynamical information on the melt can be understood in terms of the single parameter $N_e$.
Although, in general, estimating $N_e$ is a challenging problem~\cite{Lin1987,KavassalisNoolandiPRL}, considerable progress has been made (at least in numerical simulations) in terms of {\it primitive path analysis}~\cite{everaers2004rheology,kroger2005shortest,Likhtman2014} (PPA): by exploiting the simple yet ingenious idea~\cite{DoiEdwardsBook} that linear chains can be ``coarse-grained'' down to their minimal path without violating the topological constraints, PPA provides an intuitive understanding of the microscopic nature of entanglements.

Alternatively, polymeric entanglements may be also modeled as {\it physical links} between chains~\cite{Edwards1967a,Edwards1967b,Edwards1968,Iwata1989,Everaers1996,Lang2001,Likhtman2014,Caraglio2017,bobbili2020simulation,bonato2022topological}.
Specifically, the idea is ``to map'' the system of entangled chains to an equivalent one of randomly entangled (namely, self-knotted and linked) ring polymers and employ suitable {\it topological invariants}~\cite{Micheletti2011} in order to identify and then classify -- in a mathematically rigorous manner! -- the total amount of entanglements of the melt and connect them to the macroscopic viscoelastic behavior.

The connection between the two pictures is, however, not that straightforward: mainly, the reason is that the complete statistical-mechanical classification of a polymer melt would require an {\it infinite set}~\cite{Edwards1968,Everaers1996} of topological invariants in terms of pairs, triples, etc... of loops, not to mention that analytical theories are mathematically hard~\cite{Ferrari2004} and their applicability to dense systems is limited.

\begin{figure}
\includegraphics[width=0.45\textwidth]{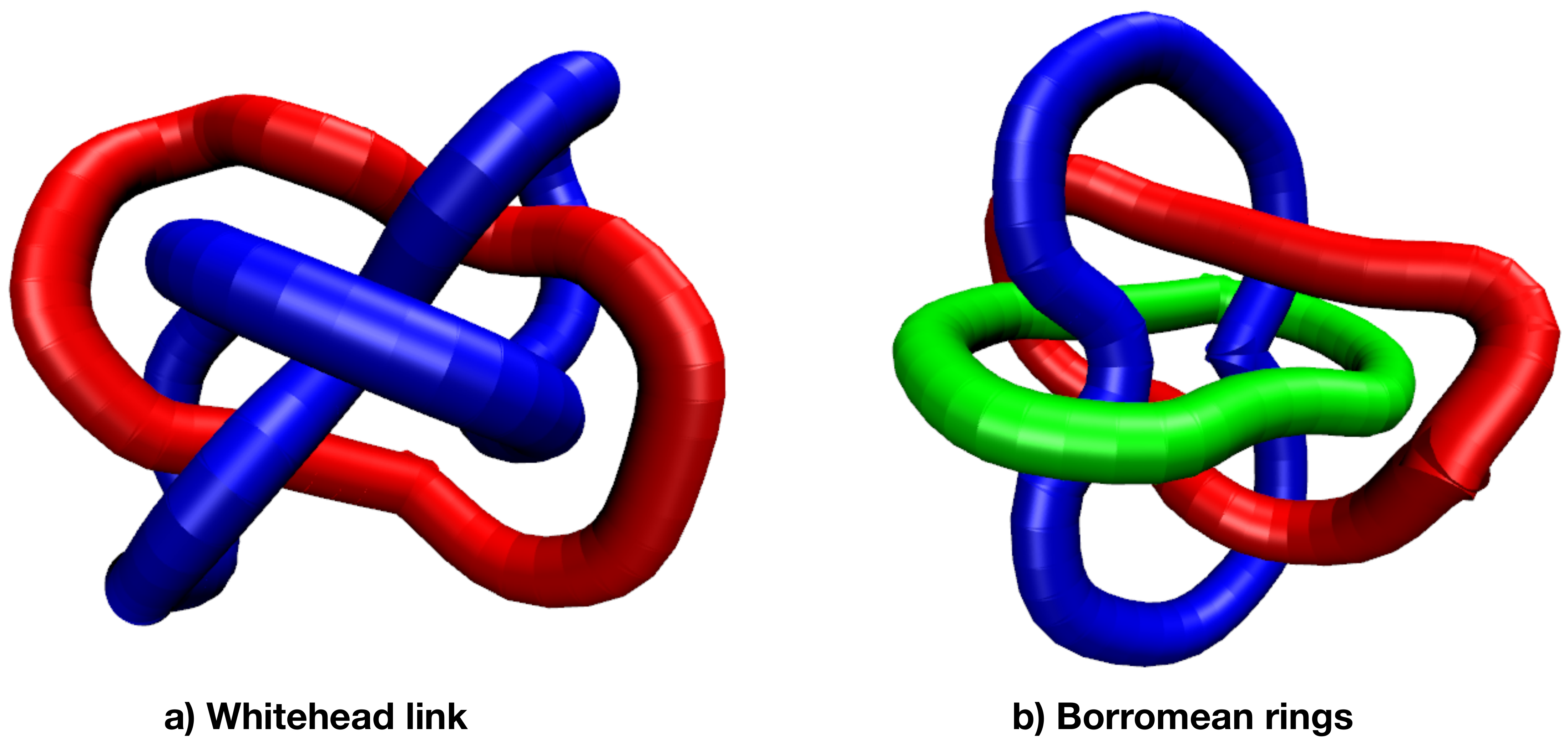}
\caption{
Examples of ring polymer structures with Gauss linking number ($\rm GLN$, see Eq.~\eqref{eq:DefineGLN}) equal to $0$.
(a)
Two rings intertwined in the Whitehead link $5^2_1$.
(b)
Three rings clustered into the Borromean conformation $6^3_2$.
Both conformations have been extracted from numerical simulations of ring polymer melts after the minimization procedure described in the text. To name the conformations here and in the rest of the text, we have used the classical nomenclature introduced in Rolfsen's book (see Sec.~\ref{sec:KnotLinkDetect}).
}
\label{fig:TypeLinks}
\end{figure}

Motivated by these considerations, in this article we rethink the problem of characterizing a melt of entangled polymer chains in terms of topological invariants and outline, in a quantitative manner, the connection between the latter and the topological entanglement length of the chains.
More specifically, we perform extensive computer simulations of {\it randomly knotted} and {\it randomly concatenated} ring polymers at dense conditions and different values of the bending stiffness of the polymer fiber as models for entangled polymer systems.

Then, inspired by PPA and by the recent work of Bobbili and Milner~\cite{bobbili2020simulation} on molecular dynamics simulations of melts of randomly linked ring polymers, we construct an algorithm for {\it contracting} the contour length of each ring in the melt to its ``primitive'' or ``minimal'' length which does not violate the topological constraints with the other rings.
The conformational properties of the primitive ring structures are thus explored at the single-ring level (knots), between any rings' pair (see the Whitehead link in Fig.~\ref{fig:TypeLinks}(a)) and between any rings' triplet (see the complex Borromean configuration in Fig.~\ref{fig:TypeLinks}(b)).
By looking at the relative abundance of these topological structures as a function of the bending stiffness of the polymers, we combine them into a proxy for the quantitative prediction of the number of entanglement lengths, $N/N_e$, of the polymers.

The paper is structured as the following:
In Section~\ref{sec:ModelMethods}, we present some technical details of the lattice polymer model, we explain the shrinking algorithm developed for the calculation of the ring minimal path and introduce the notation and the topological invariants for the characterization of knots and links and, finally, we illustrate the idea behind the Z1-algorithm used for the calculation of the entanglement length.
In Sec.~\ref{sec:Results} we present the main results of our work, while in Sec.~\ref{sec:DiscConcls} we provide some discussion and conclusions regarding the connection between knots, 2-chain and 3-chain links and the entanglement length of the polymers.
Additional figures have been included in the Supporting Information (SI) file.

\section{Model and methods}\label{sec:ModelMethods}

\subsection{Polymer model}\label{sec:PolymerModel}
Model systems of $M$ {\it concatenated} and {\it knotted} ring polymers of $N$ monomers each were prepared based on the kinetic Monte Carlo (kMC) algorithm illustrated in Refs.~\cite{ubertini2021computer,ubertini2022double}.
The polymer model, which is defined on the three-dimensional face-centered-cubic (fcc) lattice of unit step $=a$, accounts for (i) chain connectivity, (ii) bending stiffness, (iii) excluded volume and (iv) topological rearrangement of the polymer chains.
The kinetic algorithm consists of a combination of Rouse-like and reptation-like moves for chain dynamics which take advantage of a certain amount of stored contour length along the polymer filament which eases the process of chain equilibration.
This has the consequence that polymers are locally elastic, with fluctuating monomer-monomer bonds of mean length $=\langle b \rangle$ implying that the effective polymer contour length is $= N\langle b\rangle$.

\begin{table}
\begin{tabular}{cccc}
\hline
\hline
\\
\, $\kappa_{\rm bend} / (k_B T)$ \, & \, $\langle b \rangle / a$ \, & \, $\langle \cos\theta \rangle$ \, & \, $n_K$ \, \\
\hline
$0$ & $0.733$ & $0.186$ & $1.965$ \\
$1$ & $0.695$ & $0.455$ & $3.157$ \\
$2$ & $0.663$ & $0.638$ & $5.118$ \\
\hline
\hline
\end{tabular}
\caption{
Values of physical parameters for the ring polymer melts investigated in this paper.
$a$ is the unit distance of the fcc lattice and the monomer number per unit volume is $= \frac54\sqrt{2} a^{-3}$, see text and Ref.~\cite{ubertini2022double} for details.
(i) $\kappa_{\rm bend}$, bending stiffness parameter in stat. mech. thermal units $k_B T$, where $k_B$ is the Boltzmann constant and $T$ is the temperature;
(ii) $\langle b \rangle$, mean bond length~\cite{MeanBondLengthNote};
(iii) $\langle \cos\theta\rangle$, mean cosine value between two consecutive bonds along the chain~\cite{MeanBondLengthNote};
(iv) $n_K$, Kuhn length~\cite{StiffnessNote}.
}
\label{tab:PolymerModel-LengthScales}
\end{table}

Ring conformations were equilibrated through long runs 
at the average monomer number per lattice site $= \frac54 = 1.25$ or unit volume $= \frac54\sqrt{2} a^{-3}$ corresponding to melt conditions.
By modulating the Kuhn segment $n_K$ through the bending penalty Hamiltonian $\mathcal H = -\kappa_{\rm bend} \sum_{i}^{N\!\langle b\rangle /a} \cos\theta_i$, where $\kappa_{\rm bend}$ is the bending stiffness and $\theta_i$ is the angle between two consecutive bonds along the chain, it can be shown~\cite{ubertini2022double} that chains become locally stiffer: 
Table~\ref{tab:PolymerModel-LengthScales} summarizes
(i)
the mean bond length $\langle b\rangle$,
(ii)
the mean cosine value $\langle \cos\theta\rangle$ between two consecutive bonds along the chain,
(iii)
the Kuhn length $n_K$,
as a function of $\kappa_{\rm bend}$.
The simulation box of linear size $L_{\rm box}$ has periodic boundaries for the enforcement of bulky melt conditions. 
By fixing the {\it total} number of monomers to the convenient value $= 134,400$, we have that $L_{\rm box} / a = 30 \sqrt{2}$.
In this paper, we have studied polymer melts with $N\times M = (40\times3360, 80\times1680, 160\times840, 320\times420, 640\times210)$.

As illustrated in Ref.~\cite{ubertini2021computer}, we introduce random strand crossing between nearby polymer strands at the fixed rate of one per $10^4$ kMC elementary steps. 
In this way, we induce the violation of the topological constraints and obtain equilibrated melts of rings with intra-chain ({\it i.e.}, knots) and inter-chain ({\it i.e.}, links) non-trivial and randomly-generated topologies. 
By construction then, the algorithm generates rings with {\it annealed} topologies, in other words 
our ring conformations represent a thermodynamic ensemble of melts of randomly knotted and concatenated rings at the given density for different polymer lengths $N$ and bending rigidities $\kappa_{\rm bend}$.
To ensure proper system equilibration, the total computational cost of the simulations goes from $2\times10^6 \tau_{\rm MC}$ for $N=40$ and $\kappa_{\rm bend}=0$ to $7\times10^7 \tau_{\rm MC}$ for $N=640$ and $\kappa_{\rm bend}=2k_BT$.
Here, $\tau_{\rm MC}$ -- the MC ``time'' unit~\cite{ubertini2021computer,ubertini2022double} -- is equal to $N\times M$ kMC elementary steps.

Violation of topological constraints by random strand crossing induces a massive reorganization of the statistics of polymer chains.
As studied in Ref.~\cite{ubertini2021computer}, while unknotted and non-concatenated rings remain compact with asymptotic mean gyration radius following the power-law
$$
\langle R_g \rangle \sim N^{1/3} \, , 
$$
randomly knotted and randomly linked melt of rings swell as 
$$
\langle R_g \rangle \sim N^{1/2} \, , 
$$
{\it i.e.}, locally they become equivalent to melts of linear chains (see Eq.~\eqref{eq:LinChainsRg} and Fig.~S1 in SI). 
Furthermore, the distinctive anti-correlation of the bond-vector correlation function,
\begin{equation}\label{eq:BondVectorCF}
c(n) = \frac{\langle \vec t(n') \cdot \vec t(n+n') \rangle}{\langle \vec t(n')^2 \rangle} \, ,
\end{equation}
as a function of the effective monomer length separation, $n$, along the chain reported~\cite{RosaEveraersPRL2014,ubertini2022double} in melts of unknotted and non-concatenated rings disappears in randomly linked systems (see Fig.~S2 in SI), 
whose behavior is close to the one for linear chains (see dashed lines).
Overall, we may conclude that randomly linked rings reproduce the essential features of entangled linear polymer chains in melt.
In the next, we will use these systems to investigate the microscopic nature of entanglements by means of the rigorous language of topological invariants.

\subsection{Algorithmic pipeline to rings minimal paths}\label{sec:RingMinPathAlgorithm}
In order to detect and classify topological interactions in equilibrated melts of entangled rings, we introduce a simple ``shrinking'' algorithm which takes explicit advantage of the presence of stored lengths along the contour length of each chain. 
Specifically, the algorithm consists in iterating the following steps: 
\begin{enumerate}
\item
We remove away all the stored lengths from the polymers.
Of course, this excision process leads to a reduction of the total contour length of each chain.
Notice that -- by construction -- this does not lead to violations of the topological constraints, neither intra-chain ones (such as knots, for instance) nor between different chains ({\it i.e.}, links).
\item 
After the excision, we perform a short MC run (of the order of $10-100 \tau_{\rm MC}$) under global preservation of topological constraints ({\it i.e.}, without strand crossing).
In general, this step leads to formation of new units of stored length which, in turns, will be removed by the next implementation of step (1), and so on.
\end{enumerate}
The procedure stops when the number of monomers of each shrinking chain has not changed for $300$ consecutive iterations: in this case, we assume that each chain has reached its {\it minimal} shape.
To validate the algorithm, we have tested it first on the ``trivial'' case of unknotted and non-concatenated ring polymers in melt.
We have thus verified that shape minimization of rings taken one by one or simultaneous application of the procedure on the whole melt lead to what is expected based on intuition: that individual rings shrink to single points.
Then, by our algorithm, we may isolate unknotted and non-concatenated configurations from those with non-trivial topologies.

\subsection{Classification of knots and links}\label{sec:KnotLinkDetect}
Following the contour length simplification outlined in Sec.~\ref{sec:RingMinPathAlgorithm}, we have investigated 
the statistical abundance of the following topological objects: 
(i) knots in single ring polymers (Sec.~\ref{sec:Knots});
(ii) links between pairs of ring polymers ($2$-chain topological structures, Sec.~\ref{sec:2chain-Links});
(iii) links between triplets of ring polymers ($3$-chain topological structures, Sec.~\ref{sec:3chain-Links}).
We do not proceed beyond (iii) because, although in principle the procedure can be applied to even larger groups of rings, the factorial growth of possible combinations makes the analysis tediously lengthy from the computational point of view.
On the other hand, it will be shown (Sec.~\ref{sec:WholeMelt}) that this is perfectly adequate to capture the entanglement length $N_e$.

\subsubsection{Notation}\label{sec:KnotLinkNotation}
In referring to a given knot or link we follow standard convention as explained in the book by Rolfsen~\cite{Rolfsen2003KnotsAL}.
Namely, a knot or a link is defined by the symbol $K_i^p$ where:
$K$ represents the number of irreducible crossings of the knot (or the link),
$p$ is the number of rings which takes part in the topological structure ({\it e.g.}, $p=2$ for links between two rings) and $i$ is an enumerative index assigned to distinguish topologically inequivalent structures with the same $K$ and $p$.
For knots in single rings $p=1$ is tacitly assumed and, as an example, the simple trefoil knot is identified by the Rolfsen's symbol $3_1$. 

\subsubsection{Topological invariants}\label{sec:KnotDetect}
Non-trivial knots and links can be detected and hence classified by means of suitable {\it topological invariants}~\cite{Micheletti2011,orlandini2021topological}.
In this work, we resort to the method of the so called {\it Jones polynomials}~\cite{Jones1985} which assign to each knot a distinctive algebraic polynomial.
Specifically (Sec.~\ref{sec:Knots}), we use the implementation of the Jones polynomials featured in the Python package {\it Topoly}~\cite{dabrowski2021topoly} in order to recognize and categorize knots within single ring polymers and, in this way, benchmark the simplification algorithm of Sec.~\ref{sec:RingMinPathAlgorithm}. 

Moreover, and as for links alone~\cite{LinksDetectionNote}, we also consider the simpler Gauss linking number (GLN): 
\begin{equation}\label{eq:DefineGLN}
{\rm GLN} \equiv \frac1{4\pi} \oint_{{\mathcal C}_1} \oint_{{\mathcal C}_2} \frac{(\vec r_2 - \vec r_1) \cdot (d{\vec r}_2 \wedge d{\vec r}_1)}{|\vec r_2 - \vec r_1|^3} \, ,
\end{equation}
which gives the number of times two closed loops $\mathcal C_1$ and $\mathcal C_2$, parametrized respectively by coordinates $\vec r_1$ and $\vec r_2$, wind around each other. 
While intuitive and easier to compute with respect to the Jones polynomials, the $\rm GLN$ has nonetheless severe limitations~\cite{orlandini2021topological}.
It is in fact widely known that, while ${\rm GLN} \neq 0$ means that the two rings are linked, the opposite ($\rm GLN = 0$) is not necessarily true.
Take for instance the example shown in Fig.~\ref{fig:TypeLinks}(a), {\it i.e.} the so called Whitehead link $5^2_1$, constituted by two irreducibly linked rings and yet ${\rm GLN}=0$. 
On top of that, one may imagine even more complex situations such as the one displayed in Fig.~\ref{fig:TypeLinks}(b) (the so called Borromean conformation $6^3_2$) where $3$ rings, which are two-by-two non-concatenated, are irreducibly linked: such structures are, obviously, also not detected by Eq.~\eqref{eq:DefineGLN}.
In the course of the paper (Sec.~\ref{sec:Results}), we will show how these structures (which elude Eq.~\eqref{eq:DefineGLN}) can be properly detected and, then, how to quantify their impact on the entanglement properties of the melt. 

\subsection{Calculation of the entanglement length}\label{sec:ExtractLe}
By following the approach by Bobbili and Milner~\cite{bobbili2020simulation} for molecular dynamics simulations of a melt of seemingly shrunk and randomly linked ring polymers, we estimate $N_e$ by applying the recent version (Z1+~\cite{Kroger2023}) of the Z1-algorithm~\cite{kroger2005shortest,shanbhag2007primitive,karayiannis2009combined,hoy2009topological}.
The Z1-algorithm consists in the implementation of a series of geometrical operations which transform the entangled polymer chains in a collection of straight segments which are sharply bent at the entanglement points, then one may estimate $N_e$ as the average length of these straight segments. 
In particular, the Z1+ version takes explicitly into account the role of chain self-entanglements (knots) during the determination of $N_e$. The effects of it will be discussed in Sec.~\ref{sec:WholeMelt}.

\section{Results}\label{sec:Results}
In the next, we will describe results concerning the appearances of knots (Sec.~\ref{sec:Knots}) and links (Secs.~\ref{sec:2chain-Links} and~\ref{sec:3chain-Links}) in melts of entangled randomly linked rings of different chain length and bending stiffness.
Then (Sec.~\ref{sec:WholeMelt}), we will show how to establish a direct connection between the topology of links and the entanglement length of the chains.
While we have considered different chain lengths (Sec.~\ref{sec:PolymerModel}), covering the full crossover from loosely to strongly interpenetrating polymers, for brevity we will present many results only for the most representative and longest chains with $N=640$.

\subsection{1-chain topological structures, knots}\label{sec:Knots}
\begin{figure*}[t]
$$
\begin{array}{cc} 
\includegraphics[width=0.45\textwidth]{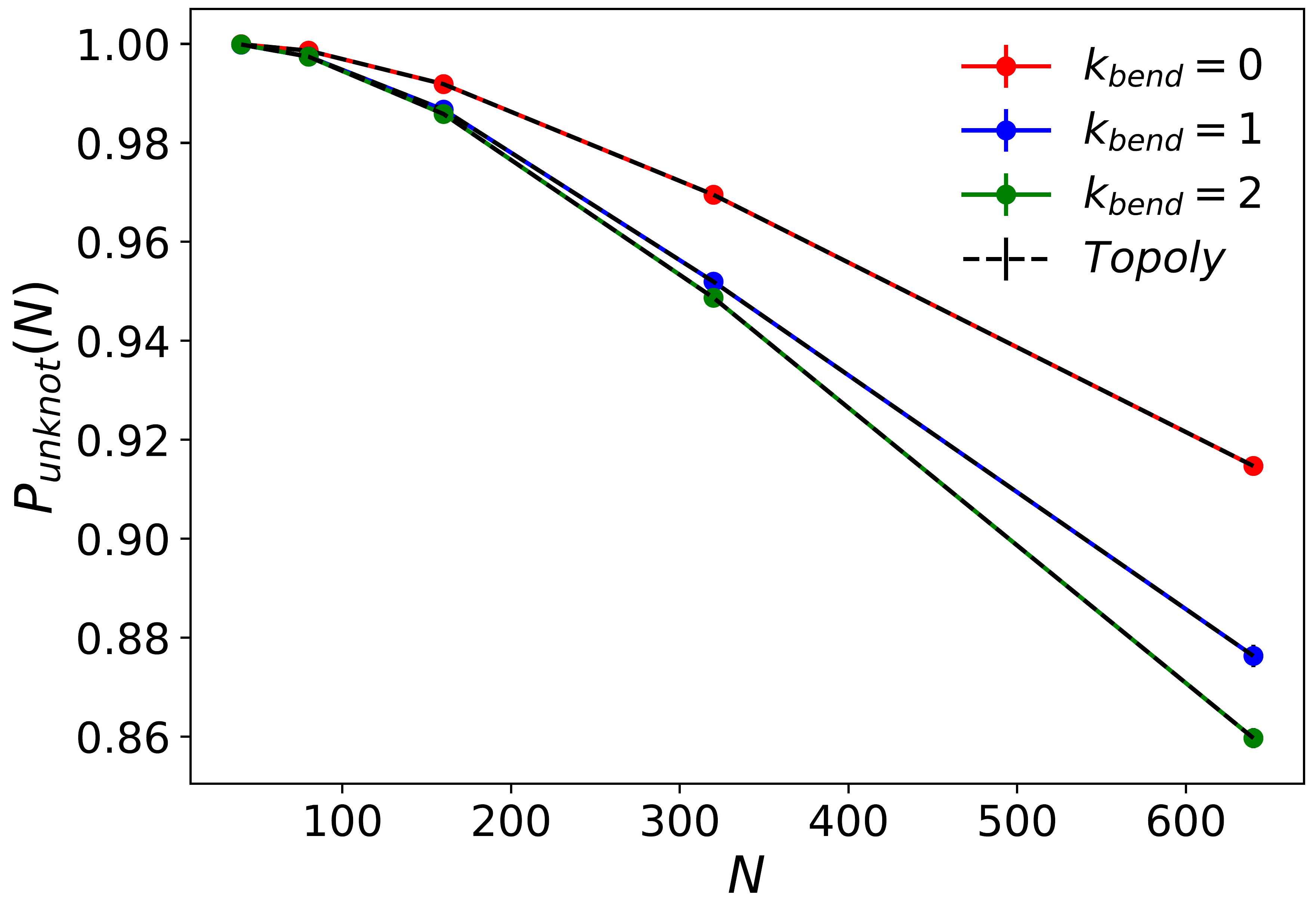} & \includegraphics[width=0.48\textwidth]{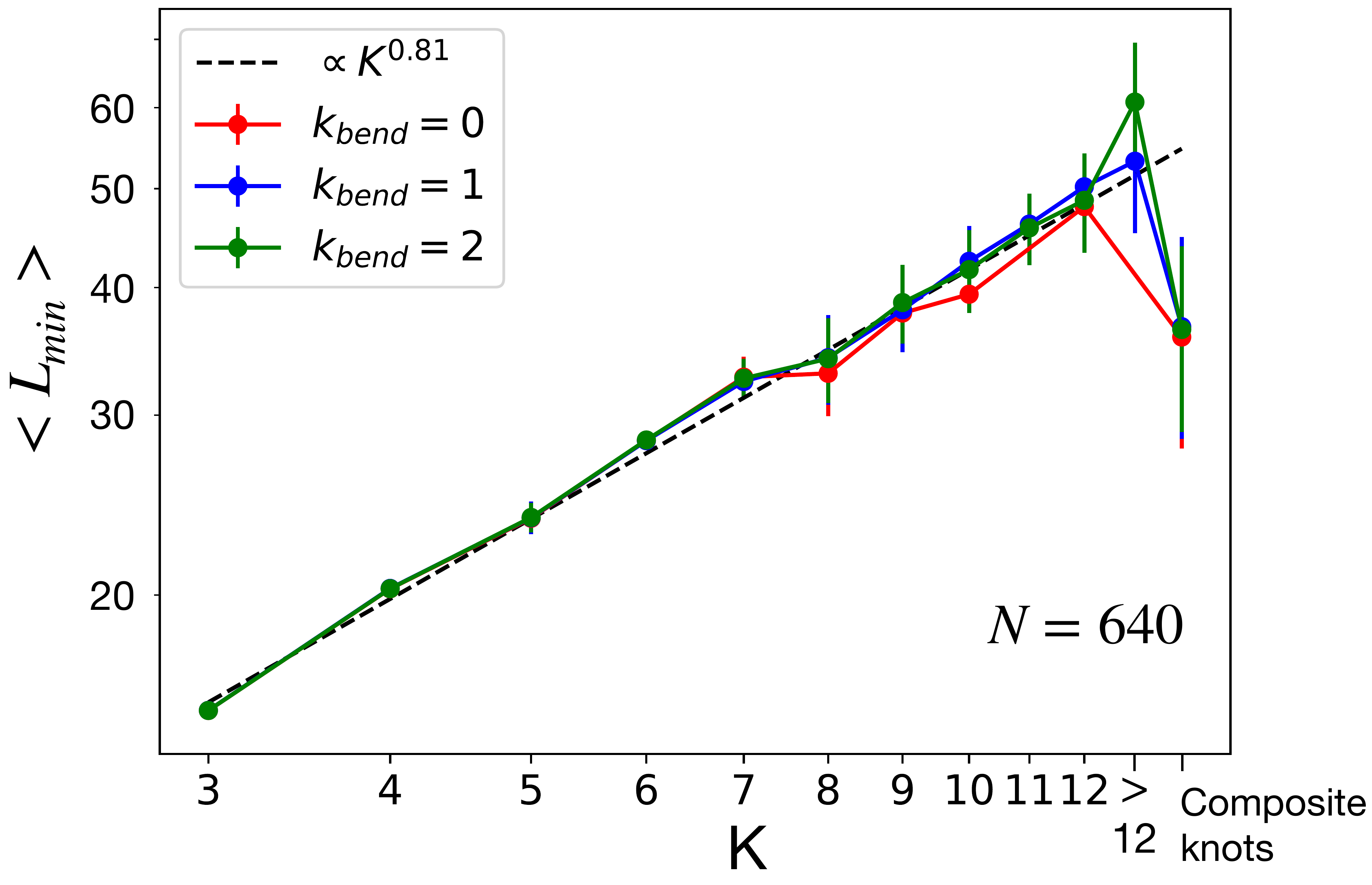}
\end{array}
$$
\caption{
(Left)
$P_{\rm unknot}$, probability that a ring is unknot as a function of the number of monomers, $N$, and for different bending stiffness, $\kappa_{\rm bend}$.
The shrinking algorithm (solid lines) and {\it Topoly} (dashed lines) are in perfect agreement. 
(Right)
$\langle L_{\rm min} \rangle$, average minimal contour length of rings with $N=640$ monomers as a function of the knot crossing number, $K$, and for different bending stiffness, $\kappa_{\rm bend}$.
Each error bar corresponds to the standard deviation calculated for the ring population at the respective crossing number $K$.
The data are well described by the simple power-law behavior $\sim K^{0.81}$ (dashed line).
The generic label ``$>12$'' follows from the fact that {\it Topoly} is unable~\cite{dabrowski2021topoly} to recognize properly knots with $>12$ crossings.
}
\label{fig:Knots}
\end{figure*}

First, we have applied our algorithm (Sec.~\ref{sec:RingMinPathAlgorithm}) to detect knots in single rings and, to prove its reliability, we have applied the {\it Topoly} tool (Sec.~\ref{sec:KnotDetect}) to the simplified ring shape in order to classify the relative knot type.
As a result, we have always found a non-trivial Jones polynomial in correspondence of those rings which do not shrink to a point, in other words the shrinking algorithm recovers knots successfully and map one-to-one to the results obtained by {\it Topoly}, see Fig.~\ref{fig:Knots} (l.h.s. panel) for the probability $P_{\rm unknot}$ that a ring is unknot as a function of the monomers number $N$ and at different bending stiffness $\kappa_{\rm bend}$. 
Overall $P_{\rm unknot}$ is always a decreasing function of the polymer length $N$, a result in line~\cite{Sumners1988,Rensburg1990} with other generic polymer models.
At the same time, for fixed $N$, $P_{\rm unknot}$ decreases as a function of $\kappa_{\rm bend}$ or stiffer rings are more likely to form knots with respect to more bendable ones and this difference appears growing with $N$: this feature seems also quite general having been reported recently~\cite{coronel2017non} in the context of computer simulations of isolated semiflexible ring polymers. 
Notice, however, that the probability to observe a knot remains small (for $\kappa_{\rm bend}/(k_BT)=2$ and $N=640$, this is only $=1-P_{\rm unknot} \approx 14\%$). 

While Jones polynomials (as well as any other topological invariant) inform us on the knot type ``trapped'' within the ring, by our shrinking algorithm we may also quantify the ``amount'' of topological entanglement ``stored'' within a non-trivial knot in terms of the corresponding ``minimal'' length scale: 
in particular, rings hosting ``simpler'' knots ({\it i.e.}, low-crossing knots) shrink more and occupy less primitive length in comparison to more complicate knots. 
To show this, we have computed the mean value, $\langle L_{\rm min} \rangle$, of the ring minimal contour length as a function of the crossing number $K$ characterizing the hosted knot.
In principle the ring minimal contour length is a random quantity because the shrinking procedures goes stochastically, on the other hand we see that the these fluctuations are, for each knot type, comparably small (Fig.~S3 in SI), 
{\it i.e.} the minimization procedure converges to a well defined minimal shape. 
Notably $\langle L_{\rm min}\rangle$ is a genuine topological signature, it is almost insensitive to the bending stiffness $\kappa_{\rm bend}$ (see Fig.~\ref{fig:Knots} (r.h.s. panel)) and it grows with the characteristic power-law $K^\alpha$ with $\alpha \simeq 0.81$ (dashed line).
Interestingly, the same power-law behaviour in relation to the scaling of the minimal rope length required to tie a non-trivial knot into a flexible rope has been reported recently~\cite{klotz2021ropelength}: we conclude then that, for a given knotted ring, our minimization algorithm converges to the corresponding minimal knot structure.
Moreover, and again in agreement with~\cite{klotz2021ropelength}, we find that the so called {\it alternating} knots, namely knots where crossings alternate under/over when moving along the filament, display bigger $\langle L_{\rm min} \rangle$ and are less frequently seen (Figs.~S3 and~S4 in SI, 
respectively, for $K\geq 8$ only~\cite{AlternatingKnotsNote}) than the {\it non-alternating} ones for the same number of crossings. 

\subsection{2-chain topological structures, links}\label{sec:2chain-Links}
After having investigated the amount of knots, we turn our attention 
to the topological interactions between {\it pairs} of rings. 
To this purpose, we have devised the following way to distinguish between those links which have Gauss linking number (Eq.~\eqref{eq:DefineGLN}) ${\rm GLN} \neq 0$ and links with ${\rm GLN} = 0$ (such as the Whitehead link, see Fig.~\ref{fig:TypeLinks}(a)).
A link between two closed chains with ${\rm GLN} = 0$ can be unlinked by performing a certain number of crossings between strands of the {\it same} chain, while the ones with ${\rm GLN} \neq 0$ can not be simplified and would remain linked.
According to that, we have applied the shrinking procedure to the two rings in the two distinct manners:
(i) straightforwardly as described in Sec.~\ref{sec:RingMinPathAlgorithm};
(ii) with {\it intra}-chain crossing allowed.
In this way, the {\it excess} of links between pairs of rings with ${\rm GLN} = 0$ can be measured as the ``difference'' between (i) and (ii). 
In order to test the robustness of this procedure, we have computed 
the corresponding Jones polynomial for the linked rings which display ${\rm GLN} = 0$. 
In the end, it turns out that only the pairs of rings which emerge as non-trivially linked feature non-trivial Jones polynomials as well.

\begin{figure*}[t]
\includegraphics[width=1\textwidth]{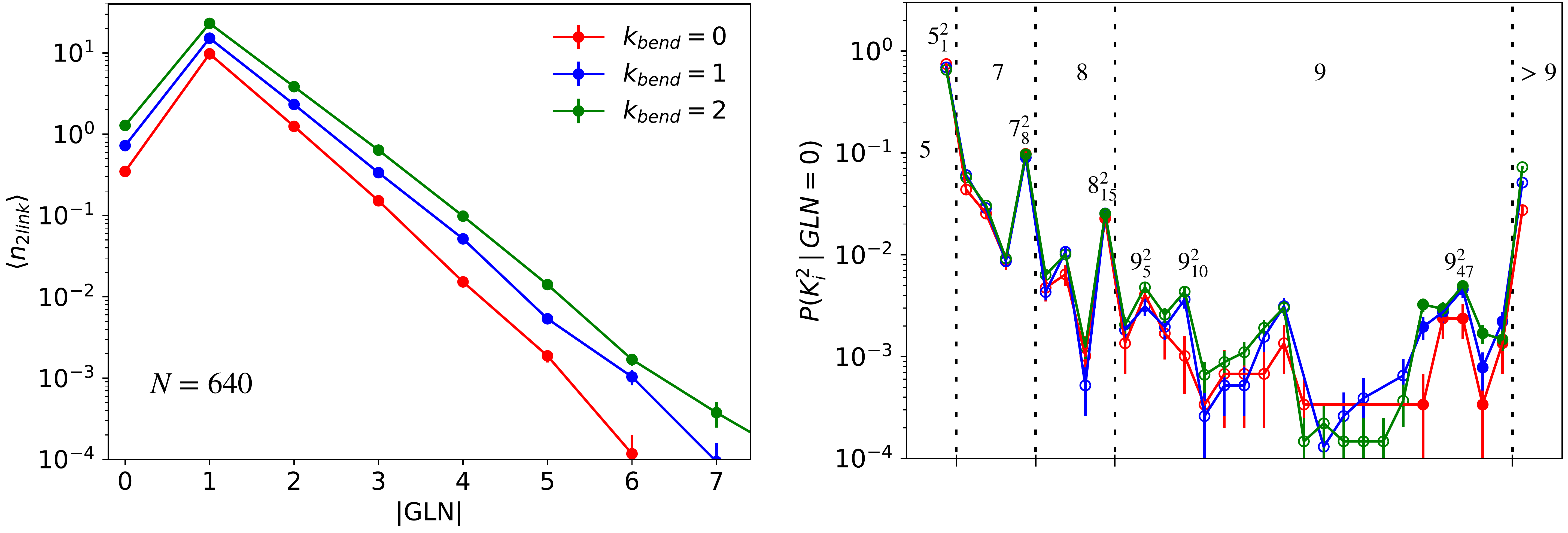} 
\caption{
(Left)
$\langle n_{2{\rm link}} (|{\rm GLN}|) \rangle $, mean number of 2-chain links per ring with absolute Gauss linking number $|{\rm GLN}|$.
(Right)
$P(K_i^2 | {\rm GLN}=0)$, fractional population of 2-chain links $K_i^2$ (termed according to the Rolfsen's convention~\cite{Rolfsen2003KnotsAL}) having ${\rm GLN} = 0$.
Here, as well as in the r.h.s. panel of Fig.~\ref{fig:Three_body} and Fig.~S4 in SI, 
error bars are estimated by assuming the formula for simple binomial statistics for the probability of observing a given link (knot, in Fig.~S4 in SI) 
type in the total population.
Empty/full circles are for alternating/non-alternating links while vertical dotted lines separate link classes with the same number of crossings.
The displayed link labels correspond to those links appearing with the highest frequency in their class of number of crossings $K$.
The generic label ``$>9$'' follows from the fact that {\it Topoly} is unable~\cite{dabrowski2021topoly} to recognize properly links with $>9$ crossings.
In both panels, data refer to rings with $N=640$ and different bending stiffness $\kappa_{\rm bend}$.
}
\label{fig:Pg}
\end{figure*}

The mean number of links per chain with absolute Gauss linking number $|{\rm GLN}|$, $n_{2{\rm link}}(|{\rm GLN}|)$, for rings with $N=640$ and different bending stiffness is shown in the l.h.s. panel of Fig.~\ref{fig:Pg} and in Fig.~S5 in SI 
for the other polymer lengths.
We find that links are mainly simple Hopf links ({\it i.e.}, $|{\rm GLN}|=1$), while links with ${\rm GLN}=0$ are rare and have frequency in between that for $|{\rm GLN}| = 2$ and $|{\rm GLN}| = 3$. More complex links follow an exponentially-decaying distribution, in agreement with~\cite{ubertini2021computer}. 
Finally, there exist many possible types of non-equivalent links for ${\rm GLN} = 0$ and we have further investigated, by the Jones polynomials, which structures emerge and their relative abundance (Fig.~\ref{fig:Pg}, r.h.s. panel).
As one may see, polymer conformations are dominated by the Whitehead link (Rolfsen's symbol: $5^{2}_1$) which, of course, is the simplest one in terms of crossings.
Nonetheless, we report a remarkably complex spectrum of link types which is very little affected by the bending stiffness of the chains. 
In particular, at number of crossings $\geq 7$, we find that the most abundant links result to be the non-alternating ones with probabilities significantly higher than the alternating ones. 
The only notable exception is for $9$ crossings where the non-alternating $9^{2}_{47}$ occurs with the same frequency of $9^{2}_5$ and $9^{2}_{10}$ which are indeed alternating:
overall, though, all these links are very rare. 

\subsection{3-chain topological structures, links}\label{sec:3chain-Links}
We consider now topological structures between ring triplets.
We point out that 3-chain links can be divided in two categories: those which can be reduced to the pair composition of 2-chain structures and those which can not or {\it irreducible}.
Those belonging to the first group are:
(a)
{\it poly(3)catenanes}, chains made of three rings in which two non-concatenated rings are connected to a common ring
and
(b)
{\it triangles}, triplets of rings which are two-by-two concatenated.
Thanks to the detection of pairwise links (Sec.~\ref{sec:2chain-Links}) their presence can be efficiently assessed.
The presence of these structures has been amply documented in melts of concatenated rings~\cite{michieletto2015kinetoplast}, in particular they can be identified -- subject to the limitations discussed in Sec.~\ref{sec:2chain-Links} -- via the summation of pairwise concatenations and the relative ${\rm GLN}$. 
On the other hand, irreducible three-chain links -- which fail detection by decomposition into pairwise linkings -- can be divided further in two classes:
(c)
{\it poly(2)catenane+1-ring}, structures made of a poly(2)catenane ({\it i.e.}, a pair of concatenated rings) plus another ring which is not directly concatenated (in a pairwise manner) with any of the two's, 
and
(d)
{\it Brunnian} links, non-trivial links which become a set of trivial links whenever one component ring is unlinked from the others (the Borromean conformation in Fig.~\ref{fig:TypeLinks}(b) constitutes the easiest example).

\begin{figure*}[ht]
\includegraphics[width=1\textwidth]{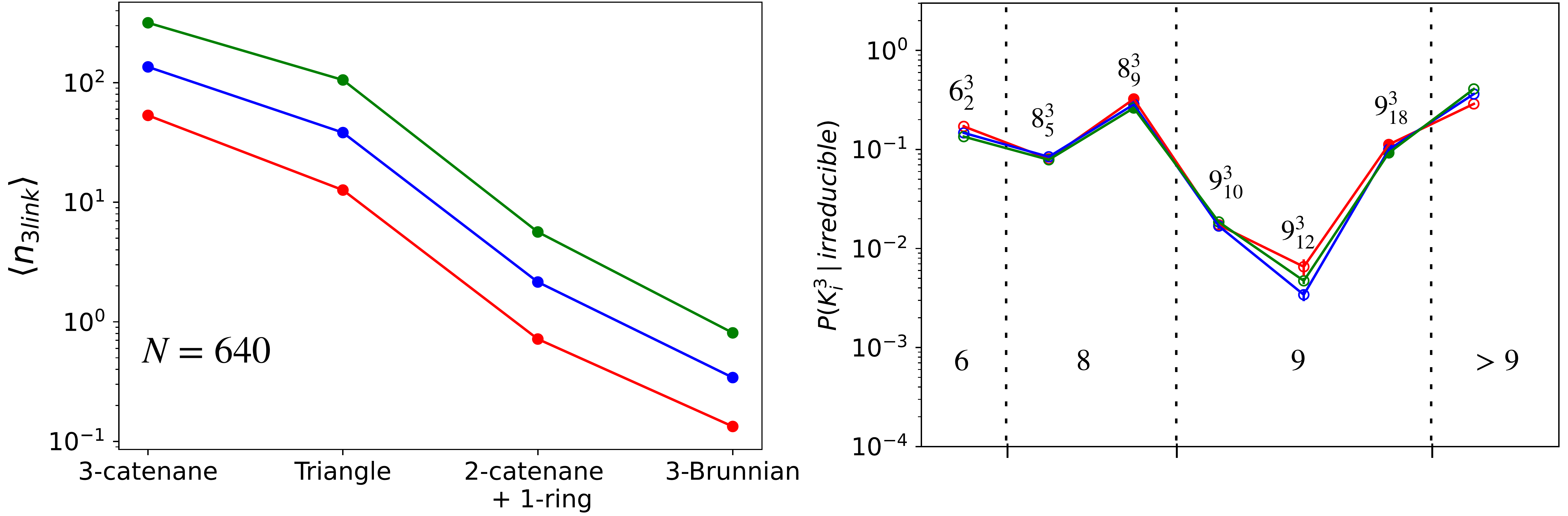} 
\caption{
(Left)
$\langle n_{3{\rm link}}\rangle$, mean number of different 3-chain structures per ring. 
(Right)
$P(K_i^3 | \mbox{irreducible})$, fractional population of 3-chain links $K_i^3$ (termed according to the Rolfsen's convention~\cite{Rolfsen2003KnotsAL}) belonging to the poly(2)catenane+1-ring and Brunnian classes (see text for details).
These are ``irreducible'' with respect to the simpler compositions of 2-chain links.
As in Fig.~\ref{fig:Pg}, empty/full circles are for alternating/non-alternating links while vertical dotted lines delimit link classes with the same number of crossings.
Similarly, the generic label ``$>9$'' follows from the fact that {\it Topoly} is unable~\cite{dabrowski2021topoly} to recognize properly links with $>9$ crossings.
In both panels, data refer to rings with $N=640$ and different bending stiffness $\kappa_{\rm bend}$.
}
\label{fig:Three_body}
\end{figure*}

In order to characterize the relative abundance of each of these structures, we have studied the mean number of different 3-chain links per ring, $\langle n_{\rm 3link}  \rangle$. 
We find (Fig.~\ref{fig:Three_body}, l.h.s. panel) that links take part maximally to poly(3)catenane and triangle structures, yet, although rarer, the other two classes appear in detectable amounts.
Notably, as for single knots and 2-chain links (l.h.s. panels of Fig.~\ref{fig:Knots} and Fig.~\ref{fig:Pg}), abundance of 3-chain structures increases with chain stiffness. 
As for the links, within the (c) and (d) classes we have analyzed the different topological inequivalent concatenated structures with {\it Topoly}.
Due to the complexity of the analyzed structures, {\it Topoly} is unable to classify them properly in about $50\%$ of the cases after $9$ crossings. 
As for the successfully determined links (Fig.~\ref{fig:Three_body}, r.h.s. panel), we get that the most abundant links are $6^3_2$ ({\it i.e.}, the Borromean rings) and $8^3_9$ (which belongs to class (c)).
Again, at fixed number of crossings, the most abundant structures are the non-alternating ones ($8^3_5$, $9^3_{10}$ and $9^3_{12}$ are all alternating), thus highlighting the preference towards non-alternating linked structures.

\subsection{Quantitative connection to the entanglement length $N_e$}\label{sec:WholeMelt}

%
\begin{figure}
\includegraphics[width=0.45\textwidth]{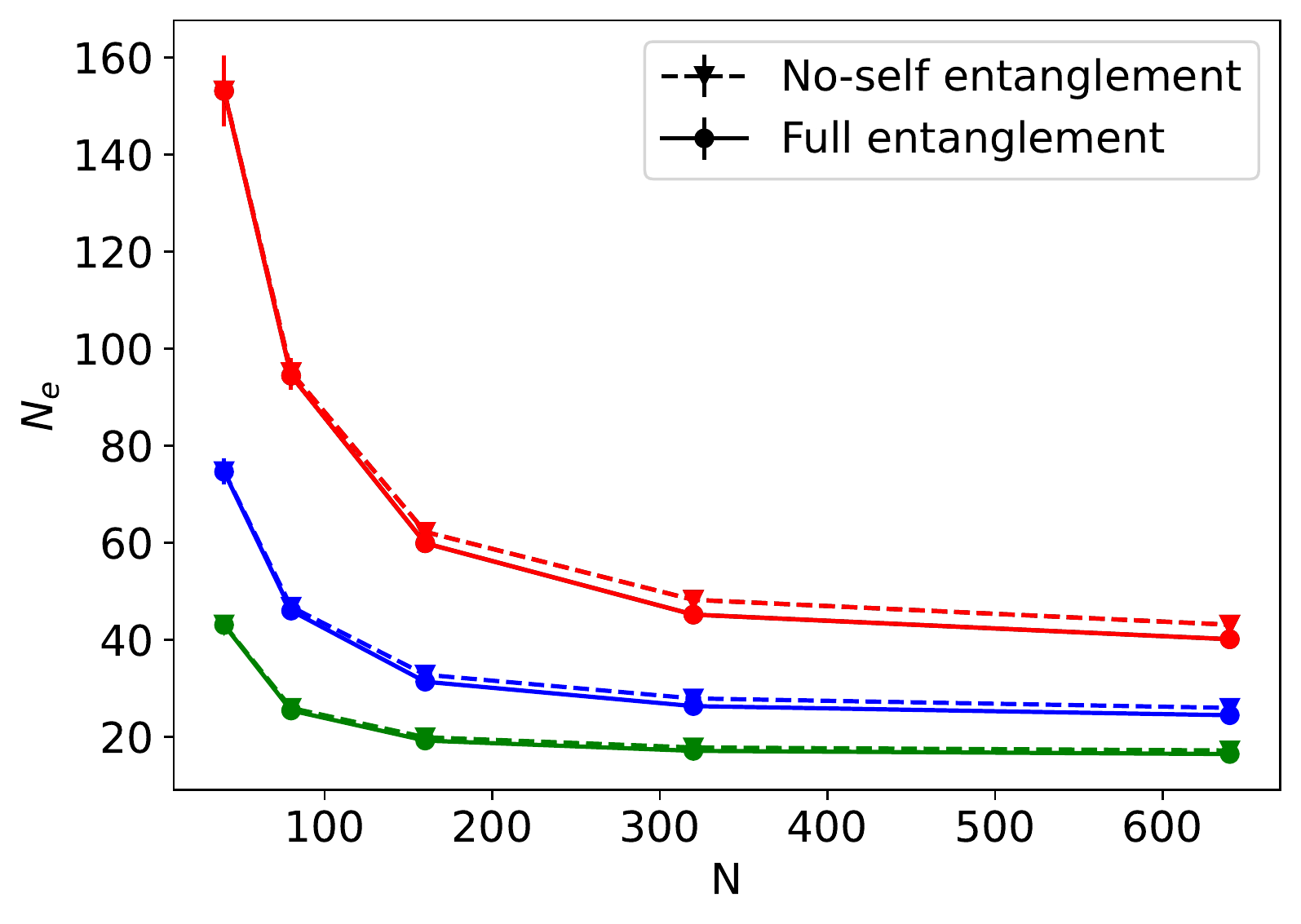}
\caption{
Entanglement length, $N_e$, as a function of the number of monomers per chain, $N$, and for different bending stiffness, $\kappa_{\rm bend}$.
Solid and dashed lines are, respectively, after including/removing self-entanglements (knots) through the Z1-algorithm (see technical details in Sec.~\ref{sec:ExtractLe}).
}
\label{fig:Ne-VS-N}
\end{figure}

By applying the shrinking algorithm to the whole melt, topological interactions of any order are taken into account and, finally, we can assess their contribution to the topological entanglement length $N_e$ (Eq.~\eqref{eq:LinChainsRg}).
In general, the process of shrinking reduces the contour length of each ring inasmuch the topological constraints allow.
Thus, if a ring is unknotted and non-concatenated it will shrink to a point and it will be not taken into account since it is assumed to not contributing to the entanglement length of the chains. Conversely, the more the rings are entangled the less they will shrink.
Then we apply (see Sec.~\ref{sec:ExtractLe} for details) the Z1-algorithm~\cite{kroger2005shortest,shanbhag2007primitive,karayiannis2009combined,hoy2009topological,Kroger2023} on the shrunk structures and estimate $N_e$ by that. 
Fig.~\ref{fig:Ne-VS-N} shows the values of $N_e$ as a function of $N$ and for the different bending stiffness $\kappa_{\rm bend}$.
In all cases $N_e$ tends to an asymptotic value ($N_e = [40.(2), 24.(5), 16.(5)]$ for $N=640$ and for $\kappa_{\rm bend}/(k_BT) = 0,1,2$, respectively), while the large values of $N_e$ measured at  small $N$ is due to the fact that rings are loosely linked, in contrast at larger values of $N$ rings result to be concatenated into a single percolating network of concatenated rings (see Fig.~S6 in SI). 

While, not surprisingly~\cite{ubertini2022double}, $N_e$ decreases as polymers become stiffer it is worth comparing these values to the ones ($N_e = [80.37(9),29.76(4),13.08(8)]$) obtained by us~\cite{ubertini2022double} by applying theoretical results based on PPA:
reasonable agreement exists for $\kappa_{\rm bend} / (k_BT) = 1,2$ while for $\kappa_{\rm bend} / (k_BT) = 0$ the new value is about a factor of $2$ off.
Interestingly, in Ref.~\cite{hoy2009topological} it has been shown that different ways of estimating $N_e$ may indeed lead to quite different results. 
While, a priori, we did not expect the same results for the two methodologies, it is unclear where the big discrepancy for the more flexible rings may come from.
Certainly (compare solid and dashed lines in Fig.~\ref{fig:Ne-VS-N}) self-entanglements ({\it i.e.}, knots) do not play a sensitive role, in agreement with the result (Sec.~\ref{sec:Knots}) that only a small fraction of the rings ($\approx 10$\%) display knots.
Overall, the difference in the entanglement length looks in agreement with earlier findings by us~\cite{ubertini2021computer} where it was shown that if the rate of strand crossing was not fast enough the system of dynamically concatenated rings may be actually slowed down with respect to the same system of unknotted and non-concatenated rings.

\begin{figure*}
\includegraphics[width=1\textwidth]{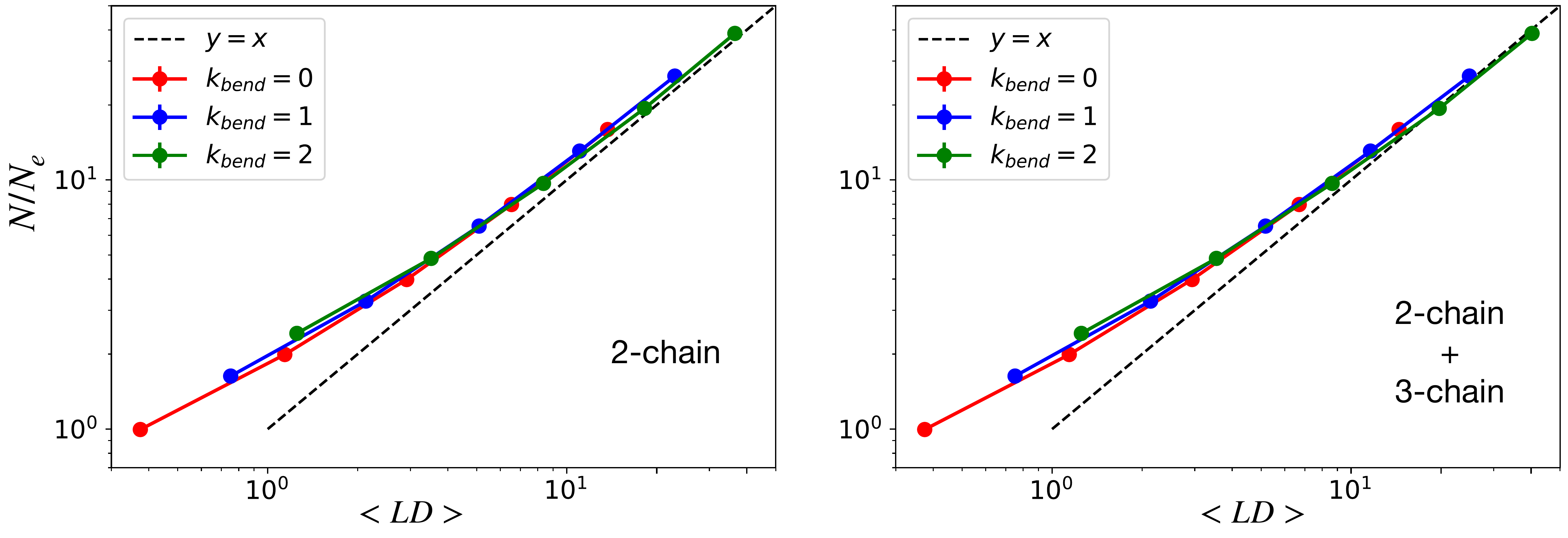}
\caption{
Number of entanglements per ring, $N/N_e$, as a function of the mean linking degree, $\langle {\rm LD} \rangle$, computed (see Eq.~\eqref{eq:MeanLD}) by taking into account the contribution from 2-chain links solely (left) and after including (right) also the contribution of 3-chain links.
}
\label{fig:Ne}
\end{figure*}

Finally we show how to connect, in a quantitative manner, $N_e$ to the linking properties of the rings (see Secs.~\ref{sec:2chain-Links} and~\ref{sec:3chain-Links}).
To this purpose, we define the ring mean linking degree $\langle {\rm LD} \rangle$ as: 
\begin{equation}\label{eq:MeanLD}
\langle {\rm LD} \rangle = \frac1M \sum_{i=1}^M \sum_{j=1}^M \chi_{ij} \, C_{ij} \, ,
\end{equation}
where each sum runs over the total number of chains ($M$, see Sec.~\ref{sec:PolymerModel}) in the melt.
$C_{ij}$ is the $M\times M$ matrix expressing the concatenation status between rings $i$ and $j$, it is defined as
\begin{equation}\label{eq:Cij-definition}
C_{ij} = \left\{ \begin{array}{cl} 0 \, , & \mbox{if $i=j$ } \\ \\ \\ 1 \, , & \mbox{if $i\neq j$ {\it and} form a 2-chain} \\ {} & \mbox{or a 3-chain irreducible link} \\ \\ 0 \, , & \mbox{otherwise} \end{array} \right.
\end{equation}
while the ``weight'' factor $\chi_{ij} = \frac{K}2$ or $\chi_{ij} = \frac{K}6$ for, respectively, 2- or 3-chain links and where $K$ is the number of crossings characterizing the link or, in other words, each crossing of the link contributes $1/2$ to an entanglement point.
Fig.~\ref{fig:Ne} (l.h.s. panel) show that, by only taking into account the contribution of 2-chain links and in the large-chain limit, Eq.~\eqref{eq:MeanLD} accounts remarkably well for the number of entanglements, $N/N_e$, of each chain.
Further inclusion (r.h.s. panel) of 3-chain links adds only a small contribution, otherwise it does not improve the agreement significantly.
This is probably the most important result of this work: it says that 2-chain links alone capture almost completely the nature of the entanglement length $N_e$ and that, through Eq.~\eqref{eq:MeanLD}, a true quantitative connection between them can be established.

\section{Discussion and conclusions}\label{sec:DiscConcls}
Understanding the microscopic nature of topological constraints in melts of polymer chains is a long-standing, classical~\cite{Edwards1967b,Edwards1968,Lin1987,KavassalisNoolandiPRL} problem in soft matter physics. 
In this work, we have characterized accurately the topological state of melts of randomly knotted and concatenated ring polymers used as models for (long) linear polymer systems and, then, show its relationship with the entanglement length $N_e$ of the chains which is {\it the} central quantity of any rheological theory~\cite{DeGennesBook,DoiEdwardsBook,RubinsteinColbyBook}. 

In order to accomplish the task, we have first shrunk the chains to their ``minimal shape'' by introducing a simple numerical algorithm which chops off progressively the contour length of the chains without producing any violation of the topological constraints present in the systems. After that, we have systematically carried out an analysis of rings' topology from the single-chain (knots) to 2- and 3-chain (links) levels. 

By using the Jones polynomials as suitable topological invariants, we have characterized the topological spectrum as a function of the bending stiffness of the chains by finding, in particular, that stiffer rings are more knotted and more concatenated with respect to more flexible ones (Figs.~\ref{fig:Knots},~\ref{fig:Pg} and~\ref{fig:Three_body}).
We have also found that, quite systematically, for both knots and links non-alternating structures are more likely to be present with respect to the alternating ones (at the same topological complexity). 
By applying the Z1-algorithm on the shrunk structures, we have computed the entanglement length $N_e$ of the melts for the different stiffnesses and found that chain self-entanglements (knots) do not play a significant role on $N_e$ (Fig.~\ref{fig:Ne-VS-N}) in fair agreement with the fact that rings are rarely knotted (Fig.~\ref{fig:Knots}).
Most importantly, we have demonstrated (Fig.~\ref{fig:Ne}) that the ring mean linking degree $\langle {\rm LD} \rangle$, which accounts for the mean number of entanglement points of each chain in the melt, is a prior for the number of entanglements $N/N_e$ which points to a non-trivial connection between the topology of the chains and the rheological entanglement of the system.
Interestingly, the quantitative matching between $\langle {\rm LD}\rangle$ and $N/N_e$ is already remarkably accurate only by including the contributions up to the simplest 2-chain linked structures suggesting that, at least for the chain lengths examined here, links of higher orders contribute negligibly. 
Overall, these findings highlight the connection between the rheological entanglements and the topological links between distinct chains acting at the microscopic level. 

To conclude, while this work is mostly focused on understanding the relation between the rheological entanglement of the melt and the microscopic topological state of its constituent chains, model conformations of randomly knotted and concatenated rings can be adopted~\cite{ubertini2021computer} to understand the mechanisms of synthesis of so called Olympic gels, namely polymer gels made of randomly linked rings like the ones now realized by using DNA and cutting restriction enzymes~\cite{krajina2018active}.
In particular, the possibility to perform fine tuning of the fiber parameters allow to foresee in great detail how one can benefit from the topological properties of the gel and design materials with certain specificities. 
For instance, a by-product of the present work concerns how the polymer length, combined with the bending stiffness of the chain, influence the topology of the resulting structure.
Depending on $\kappa_{\rm bend}$, there is a different critical $N$ for which a percolating network of concatenated rings appears (Fig.~S6 in SI), 
in particular longer and stiffer rings typically produce more robust networks. 
Moreover, depending on $N$ and $\kappa_{\rm bend}$ the networks are constituted by a complex zoo of catenation motifs: Hopf links, which are the most abundant for all considered $N$ and $\kappa_{\rm bend}$ (Fig.~\ref{fig:Pg} (l.h.s. panel) and Fig.~S5 in SI), 
some more complex links with ${\rm GLN} = 0$ ({\it e.g}, the Whitehead link) and $|{\rm GLN}| > 1$ or links involving 3-chain structures whose abundances grow with $N$ and $\kappa_{\rm bend}$ (see Fig.~\ref{fig:Pg} (r.h.s. panel) and Fig.~\ref{fig:Three_body}).
These considerations highlight the topological complexity which may arise in Olympic gels made up by strand-crossing rings as in~\cite{krajina2018active} and how topology can be fine regulated by controllable external parameters such as $N$ and $\kappa_{\rm bend}$.

{\it Acknowledgments} --
The authors are indebted with M. Kr\"oger for sharing with us the algorithm Z1+ before its official release~\cite{Kroger2023} and with P. Dabrowski-Tumanski who gave us invaluable technical advice regarding the {\it Topoly} package~\cite{dabrowski2021topoly}.
The authors also acknowledge networking support by the COST Action CA17139 (EUTOPIA).

\bibliography{biblio}

\begin{thebibliography}{43}
\bibitem [{OnC()}]{OnContourLengthUnitsNote}%
  \BibitemOpen
  \href@noop {} {}\bibinfo {note} {In this work, polymer contours are measured
  in monomer units and termed (for compactness and with abuse of language)
  ``lengths'': so, $n_K$ and $N_e$ are, respectively, the Kuhn and the
  entanglement length of the polymer. Conversion to ``true'' length is obtained
  by multiplying for the mean bond length $\langle b\rangle$.}\BibitemShut
  {Stop}%
\bibitem [{Lin()}]{LinksDetectionNote}%
  \BibitemOpen
  \href@noop {} {}\bibinfo {note} {The shrinking algorithm can be used to
  isolate structures of linked rings, yet it is easy to realize that some care
  has to be taken due to the fact that two or more shrinking rings may be stuck
  in non-trivial shapes either because they are effectively linked or because,
  although not linked, they are both non-trivially knotted. To remove such a
  possible ambiguity, we assign two rings a link if the spatial distance
  between their centers of mass is $<2 R_g$ where $R_g$ is the mean gyration
  radius of the chain. For $3$-ring link detection, this same criterion has
  been applied to each ring pair in the triplet. We also mention that -- and
  only for the purpose of the detection -- before minimization each of the
  examined structures should be placed at the center of a large box ensuring
  that no spurious entanglements emerge as a consequence of the periodic
  boundary conditions (Sec.~\ref{sec:PolymerModel}) introduced in the
  model.}\BibitemShut {Stop}%
\end{thebibliography}

\clearpage

\widetext
\clearpage
\begin{center}
\textbf{\Large Topological analysis and the recovery of entanglements in polymer melts \\ \vspace*{1.5mm} -- Supporting Information --} \\
\vspace*{5mm}
Mattia Alberto Ubertini and Angelo Rosa
\vspace*{10mm}
\end{center}
\balancecolsandclearpage

\setcounter{equation}{0}
\setcounter{figure}{0}
\setcounter{table}{0}
\setcounter{page}{1}
\setcounter{section}{0}
\makeatletter
\renewcommand{\theequation}{S\arabic{equation}}
\renewcommand{\thefigure}{S\arabic{figure}}
\renewcommand{\thetable}{S\arabic{table}}
\renewcommand{\thesection}{S\arabic{section}}

\makeatletter
\@fpsep\textheight
\makeatother

\begin{figure*}
$$
\begin{array}{ccc} 
\includegraphics[width=0.32\textwidth]{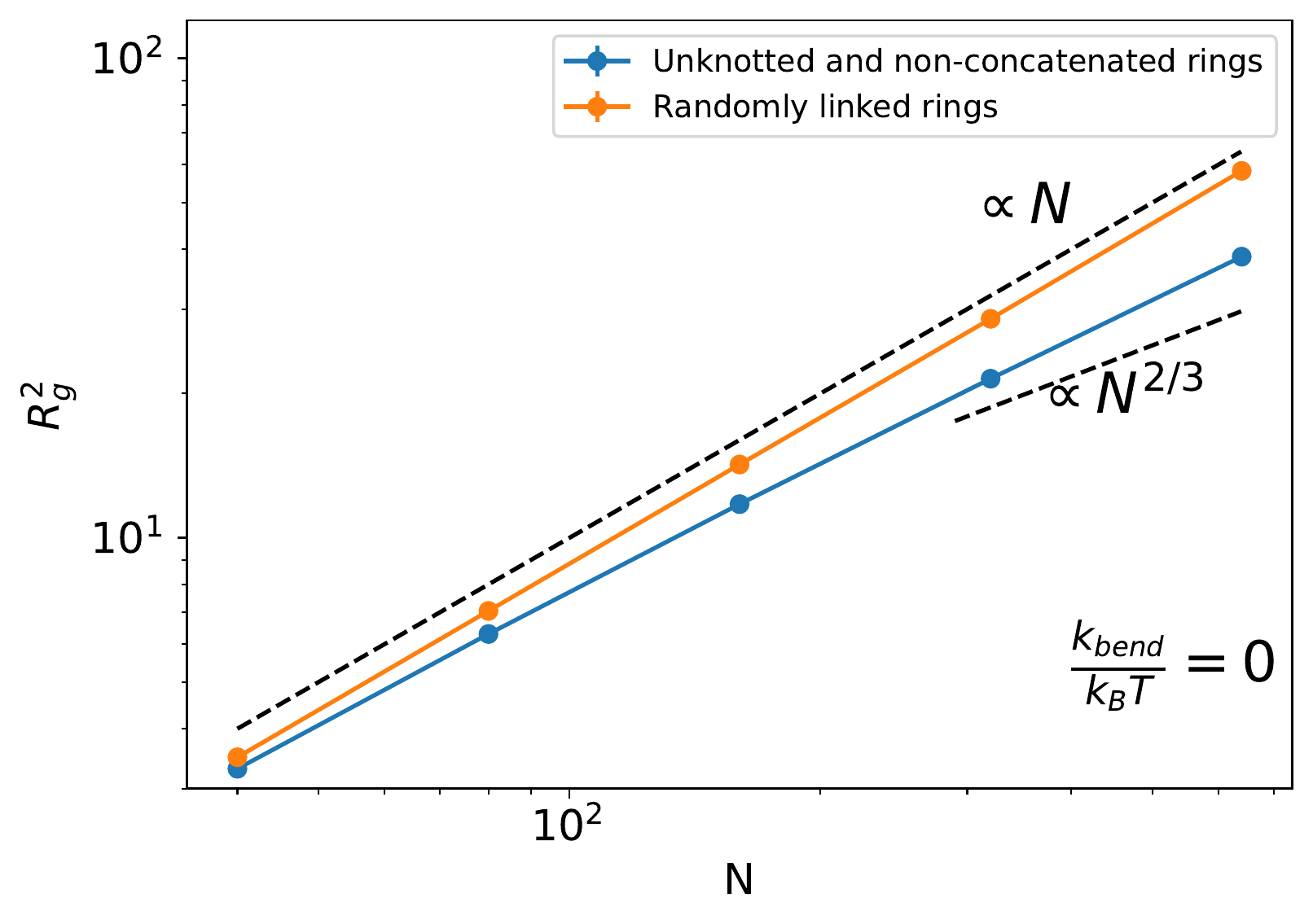} & \includegraphics[width=0.32\textwidth]{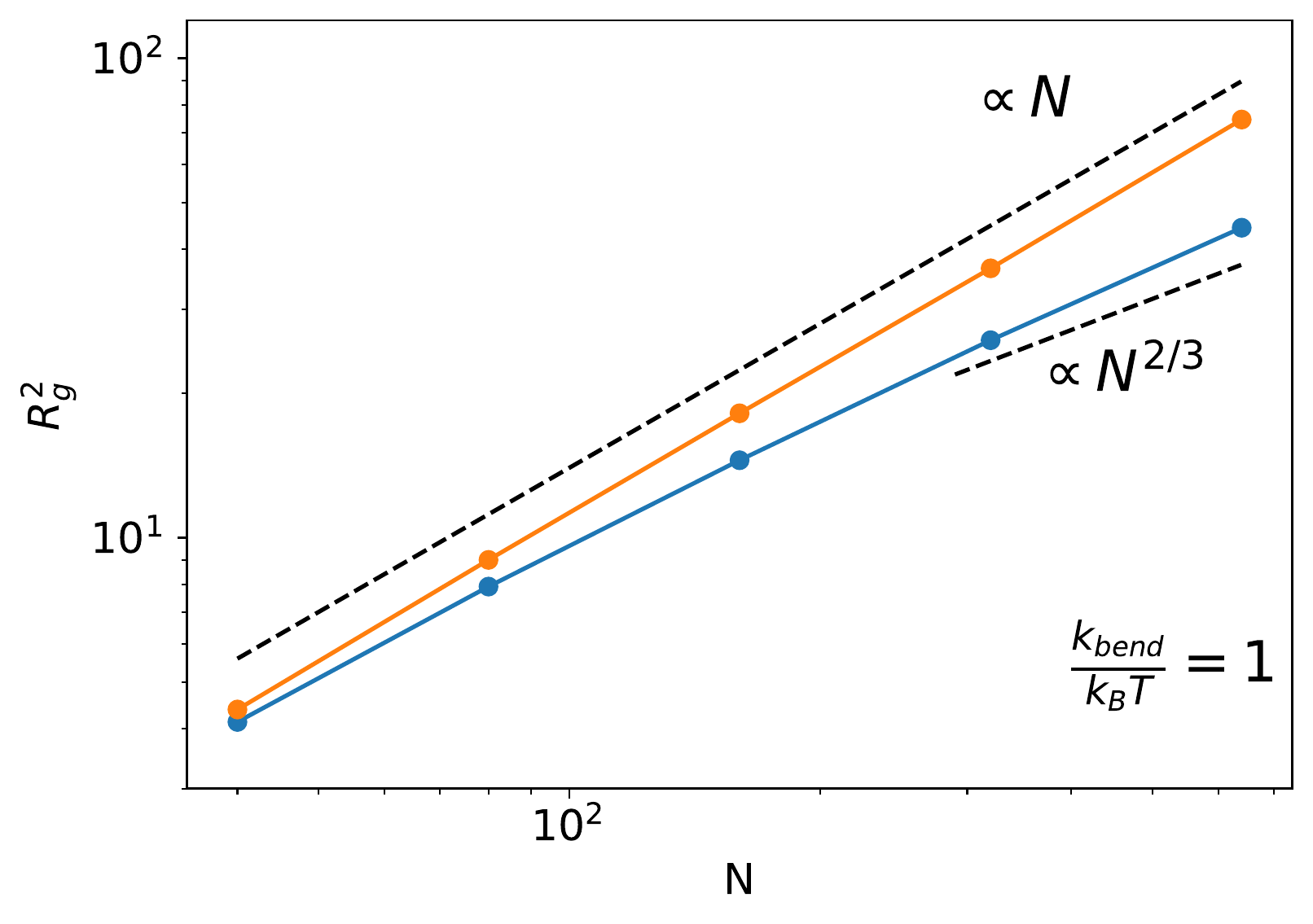} & \includegraphics[width=0.32\textwidth]{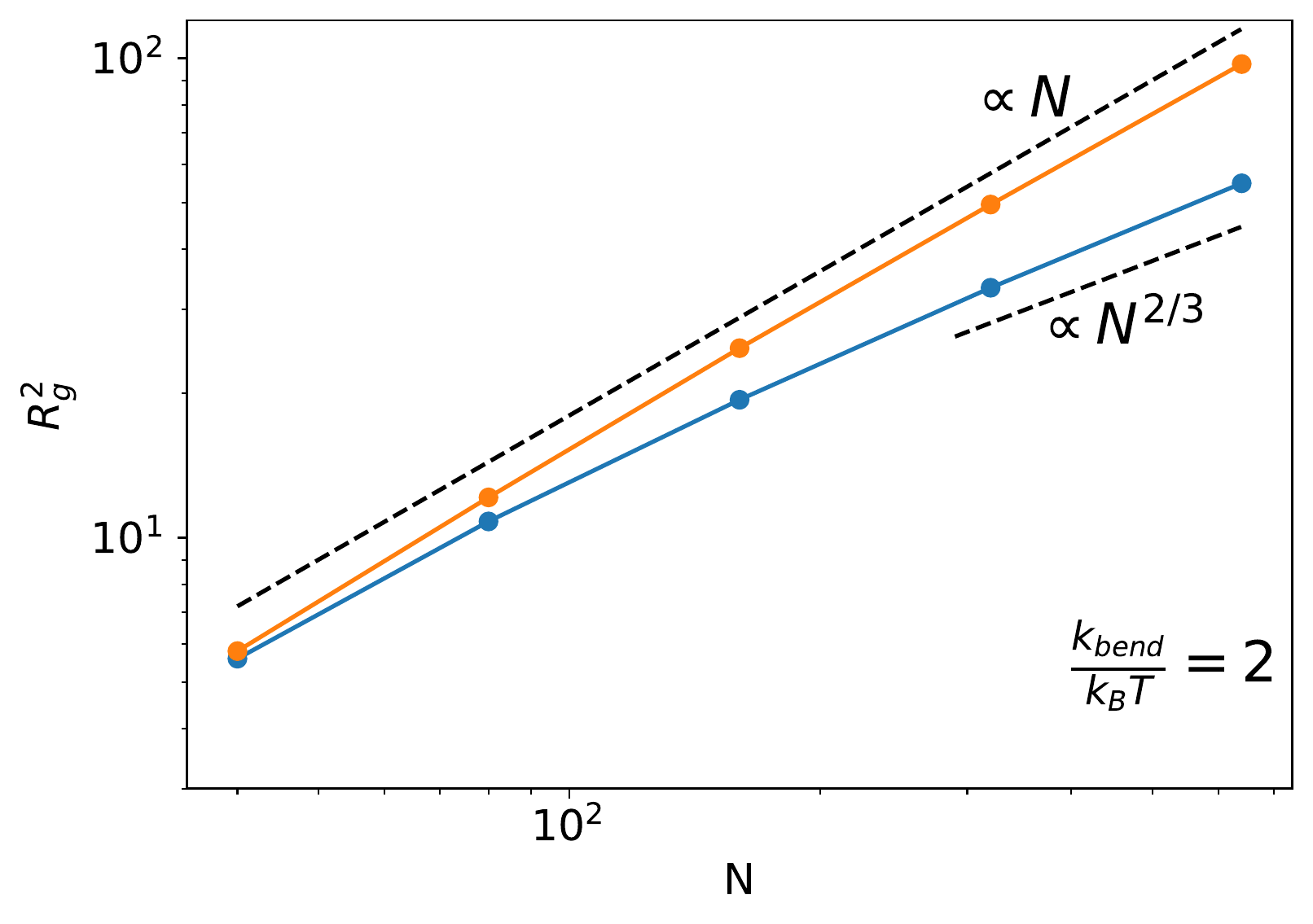}
\end{array}
$$
\caption{
Mean-square gyration radius $\langle R^2_g \rangle$ for melts of unknotted and non-concatenated rings (data from Ref.~\cite{ubertini2022double}) and randomly linked rings as a function of the total number of monomers per ring, $N$.
Panels from left to right are for bending stiffnesses $\kappa_{\rm bend} / (k_BT) =0,1,2$ (see label). 
}
\label{fig:Rg_comparison}
\end{figure*}
\begin{figure*}
$$
\begin{array}{cc} 
\, \, \, \, \, \, \, \, \, \, \, \, \, \, \, \, \, \, \mbox{(a) Unknotted and non-concatenated rings} & \, \, \, \, \, \, \, \, \, \, \, \, \, \, \, \, \, \mbox{(b) Randomly linked rings} \\
\includegraphics[width=0.45\textwidth]{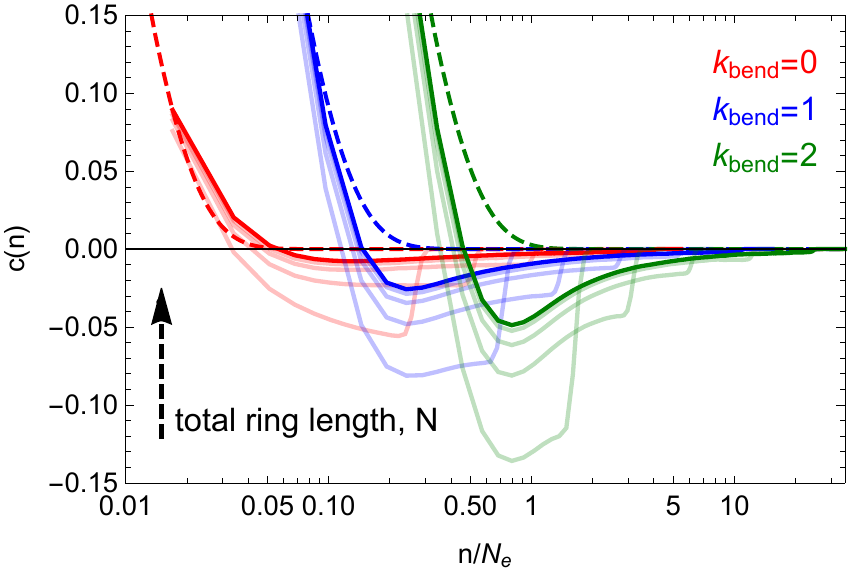} & \includegraphics[width=0.45\textwidth]{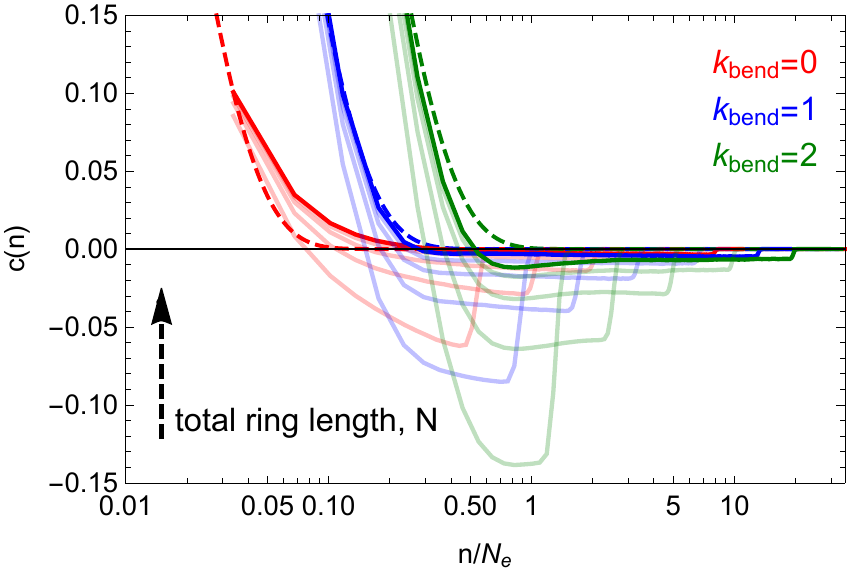}
\end{array}
$$
\caption{
Bond-vector correlation function $c(n)$ (Eq.~\eqref{eq:BondVectorCF} in the main paper) as a function of the effective monomer length, $n / N_e$, normalized with respect to the entanglement length $N_e$.
The l.h.s. and r.h.s panels are for
(a)
melts of unknotted and non-concatenated rings (data from our previous work Ref.~\cite{ubertini2022double})
and
(b)
randomly linked rings studied in this paper.
Lines of equal color are for the same chain stiffness ($\kappa_{\rm bend}$ in units of $k_BT$, see legend), full colors are for the longest rings ($N = 640$), while lines in fainter colors are for chains of shorter contour lengths (see arrow's direction).
The long-dashed lines correspond to the exponential decay typical of linear polymers with local stiffness, {\it i.e.} $c(n) = \langle \cos\theta\rangle^n$.
The values for $\langle \cos\theta \rangle$ and $N_e$ used in panels (a) and (b) are from, respectively, Ref.~\cite{ubertini2022double} and the present work (see Table~\ref{tab:PolymerModel-LengthScales} and Sec.~\ref{sec:WholeMelt} in the main paper). 
}
\label{fig:BondVectorCorrelFunct}
\end{figure*}
\begin{figure*}
\includegraphics[width=1\textwidth]{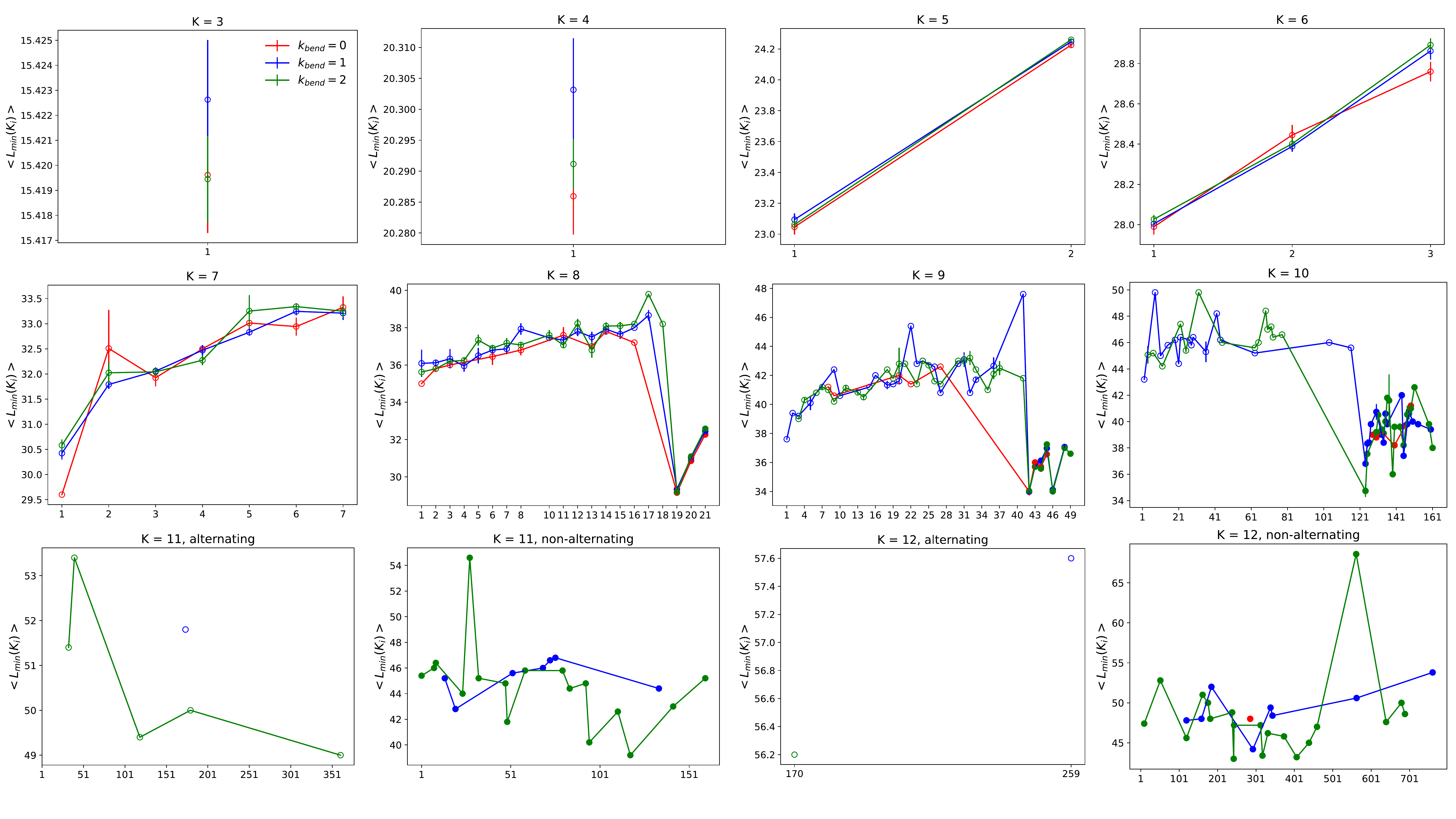}
\caption{
Average ring minimal contour length (with error bars), $\langle L_{\rm min}(K_i) \rangle$, computed for each knot type.
The labels on the $x$-axis are for each possible knot at the given $K$; for $K\geq 9$, the labels (starting from $1$) appear with regular spacing for reasons of space (except the panel ``$K=12$, alternating'' where only two knots have been detected).
As in Figs.~\ref{fig:Pg} and~\ref{fig:Three_body} in the main text, empty/full circles are for alternating/non-alternating knots.
As in the rest of the paper (see Sec.~\ref{sec:KnotLinkNotation} in the main text), knots with $K\leq 10$ crossings are named according to the Rolfsen's convention while knots with $K=11$ and $K=12$ crossings are conventionally~\cite{dabrowski2021topoly} split into alternating, $K_{{\rm a\_i}}$, and non-alternating, $K_{{\rm n\_i}}$, ones with the ordered index $i\geq 1$ in both cases.
Data with no error bars are for rare knot types, which occur only once in the generated melt conformations. 
The results shown here are for rings with $N=640$ monomers and different values of the bending stiffness, $\kappa_{\rm bend}$.
}
\label{fig:AnalysingLmin}
\end{figure*}
\begin{figure*}
\includegraphics[width=1.\textwidth]{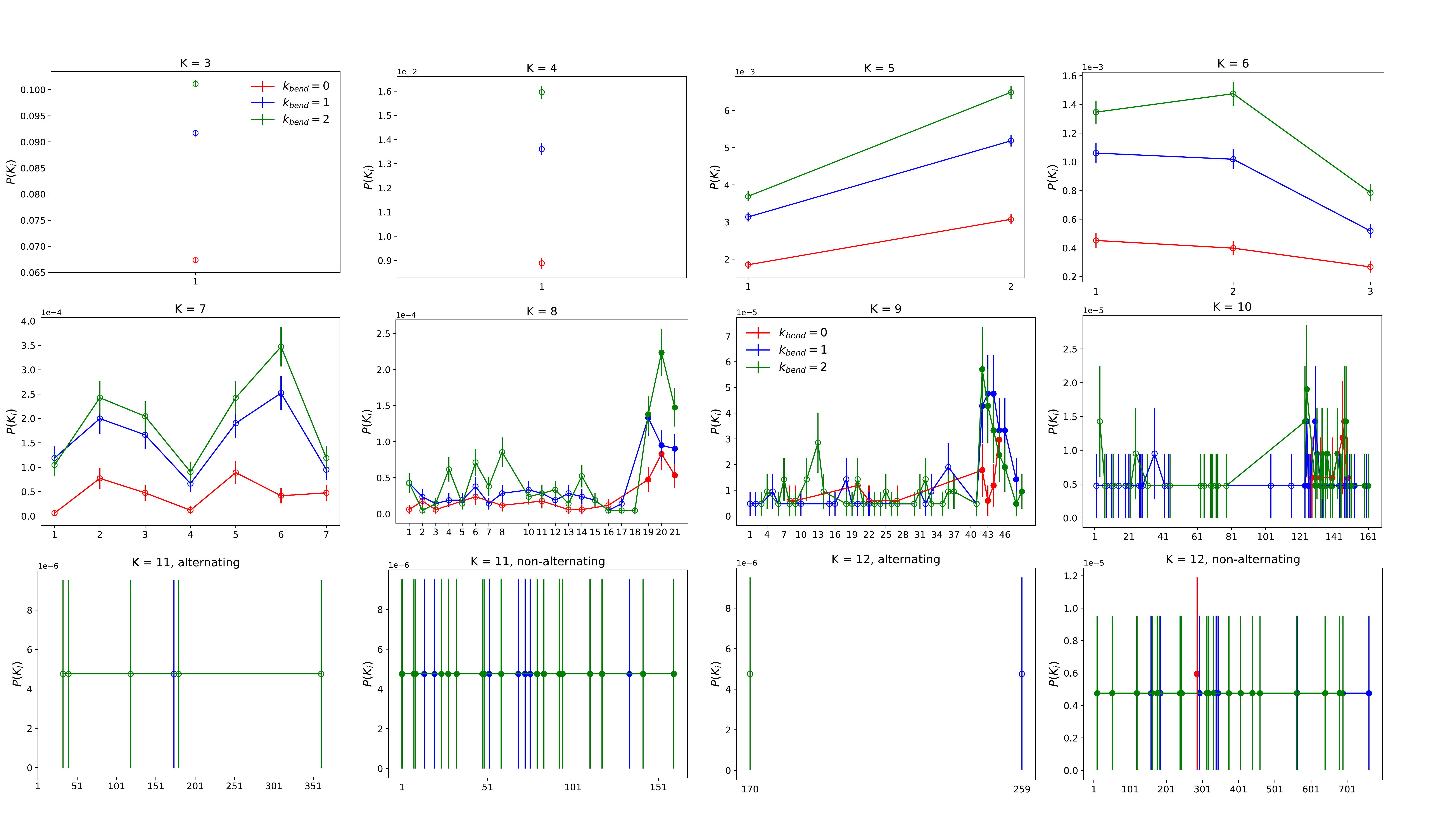}
\caption{
Fractional population (with error bars), $P(K_i)$, computed for each knot type.
Notice that the values on each $y$-axis have to be multiplied by the power-law reported on the top left corner of the corresponding panel.
Symbols, labels and notation are as in Fig.~\ref{fig:AnalysingLmin}.
Large error bars are due to the limited size of the relative sample.
}
\label{fig:KnotTypeFreq}
\end{figure*}
%


%
\begin{figure*}
\includegraphics[width=1\textwidth]{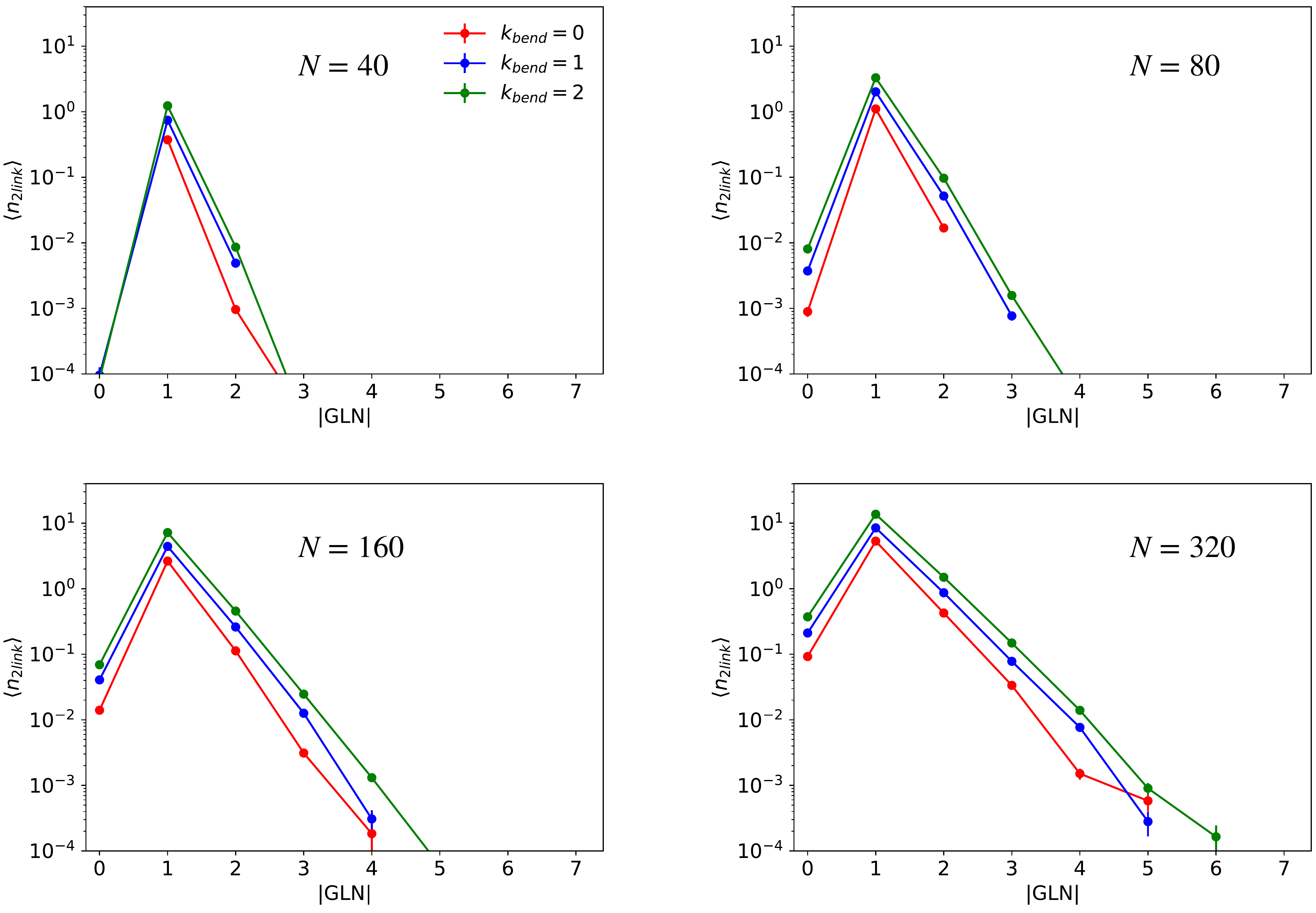}
\caption{
$\langle n_{2{\rm link}} (|{\rm GLN}|) \rangle $, mean number of 2-chain links per ring with absolute Gauss linking number $|{\rm GLN}|$.
Results for rings with $N$ monomers (to be compared to the results for $N=640$ reported in the l.h.s. panel of Fig.~\ref{fig:Pg} in the main text).
}
\label{fig:P-G_list}
\end{figure*}
\begin{figure*}
\includegraphics[width=1\textwidth]{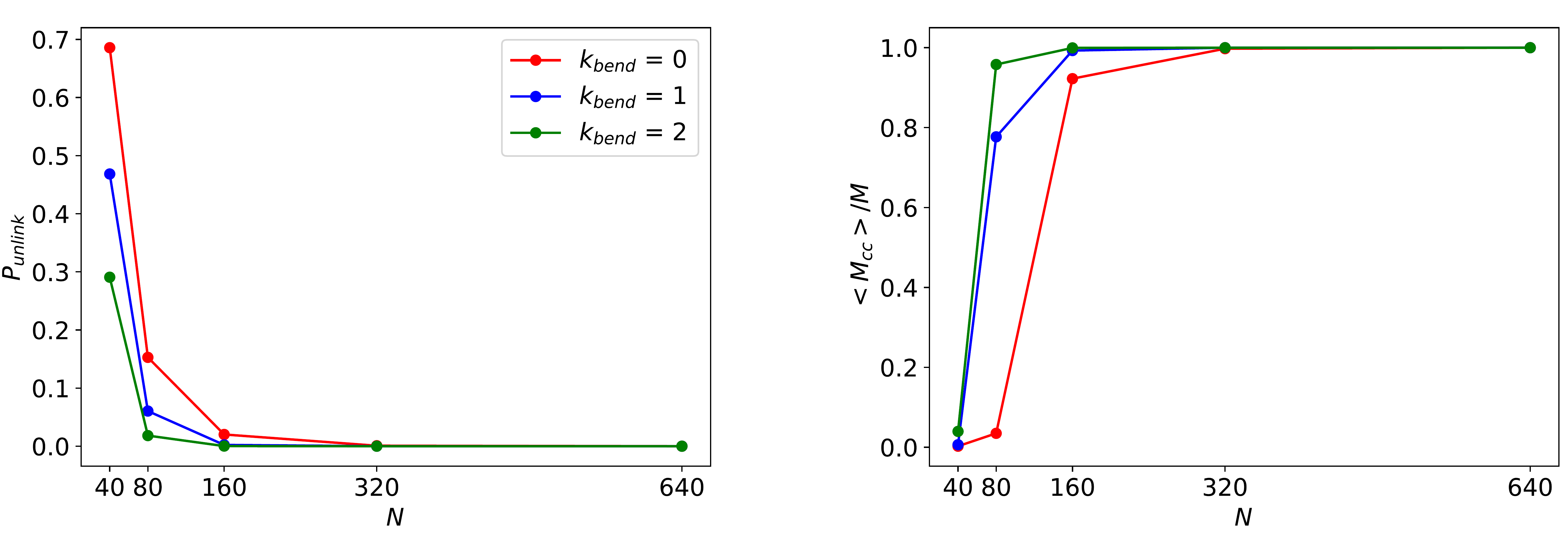}
\caption{
(Left)
Probability, $P_{\rm unlink}$, that a ring is {\it not} concatenated to any other ring of the system as a function of $N$ and $\kappa_{\rm bend}$.
(Right)
Mean fraction of rings, $\langle M_{\rm cc} \rangle / M$, belonging to the {\it largest connected component} of chains in the melt as a function of $N$ and $\kappa_{\rm bend}$.
}
\label{fig:Unlink-Giant}
\end{figure*}
%

\end{document}